\documentclass[twocolumn,iop,apj]{aastex63}
\usepackage{amsfonts,amsmath,graphicx,natbib,url,hyperref,nicefrac}
\usepackage{amssymb, verbatim, subfigure, paralist, soul, comment}

\hypersetup{colorlinks=true, urlcolor=blue, citecolor=blue}

\usepackage{color}
\newcommand{\sprout}{\texttt{Sprout}}
\newcommand{\los}{$l_0^{\mathrm{S}}\,$}
\newcommand{\lofe}{$l_0^{\mathrm{Fe}}\,$}
\newcommand{\add}[1]{\textcolor{black}{#1}}

\shorttitle{Double Detonation SNRs} 
\shortauthors{Mandal et al.}

\begin{document}

\title{Deciphering the explosion mechanism of Type Ia SNe using their remnants II: a deep dive into double detonations with SNR 0509-67.5}

\author[0000-0001-9484-1262]{Soham Mandal}

\affiliation{Department of Astronomy, University of Virginia, 530 McCormick Road, Charlottesville, VA 22904, USA}
\affiliation{Virginia Institute for Theoretical Astronomy, University of Virginia, Charlottesville, VA 22904, USA}

\author[0000-0002-9886-0839]{Parviz Ghavamian}
\affiliation{Department of Physics, Astronomy and Geosciences,
  Towson University, Towson, MD, 21252; pghavamian@towson.edu}
  
\author[0000-0002-5483-0232]{Priyam Das}
\affiliation{School of Science, The University of New South Wales, Australian Defence Force Academy, Northcott Drive, Campbell, Canberra, ACT 2600, Australia}

\author[0000-0002-5044-2988]{Ivo Rolf Seitenzahl}
\affiliation{Mathematical Sciences Institute, Australian National University, Canberra, ACT 0200, Australia}

\author[0000-0002-1856-9225]{Shazrene Mohamed}
\affiliation{Department of Astronomy, University of Virginia, 530 McCormick Road, Charlottesville, VA 22904, USA}
\affiliation{Virginia Institute for Theoretical Astronomy, University of Virginia, Charlottesville, VA 22904, USA}
\affiliation{South African Astronomical Observatory, P.O Box 9, Observatory, 7935, Cape Town, South Africa}
\affiliation{Department of Astronomy, University of Cape Town, Private Bag X3, Rondebosch, 7701, Cape Town, South Africa}
\affiliation{NITheCS National Institute for Theoretical and Computational Sciences, South Africa}

\author[0000-0002-4794-6835]{Ashley J. Ruiter}
\affiliation{School of Science, The University of New South Wales, Australian Defence Force Academy, Northcott Drive, Campbell, Canberra, ACT 2600, Australia}

\email{soham@virginia.edu}

\begin{abstract}

\add{Type Ia supernovae (SNe) occur when a white dwarf (WD) explodes via runaway thermonuclear burning. Till date, major uncertainties remain regarding the nature of the explosion mechanism and its observable signatures. In this work, we study how the `double detonation' explosion mechanism, or a helium shell detonation in a sub-Chandrasekhar WD followed by a core detonation, shapes supernova remnants (SNRs) and encodes information about the WD progenitor. We evolve a suite of double-detonation SN models to the remnant phase, up to several centuries after explosion, and measure the characteristic sizes of substructures formed in the SNR due to turbulent mixing. By comparing our models to high-resolution optical observations of the young Type Ia SNR 0509–67.5, we find that the size distribution of its small-scale substructures is consistent with a double detonation explosion mechanism and further places constraints on the carbon–oxygen core mass and helium shell mass of the WD progenitor. The observed sizes of iron-dominated and sulfur-dominated substructures in SNR 0509–67.5 indicate a progenitor core mass and a shell mass of $1M_{\odot}$ and $\gtrsim0.05M_{\odot}$, respectively.}

\end{abstract}


\section{Introduction}  \label{sec:intro}

Type Ia supernovae (SNe Type Ia) are runaway thermonuclear explosions of white dwarf (WD) stars assumed to be rich in carbon and oxygen. They constitute an important pathway to the formation of elements in the present-day universe \cite[and are the dominant source of iron-group elements; see][]{Maoz+2017ApJ}. They are regarded as crucial distance indicators in the cosmological distance ladder \citep{Kirshner2010}, owing to their famous characteristic of having absolute luminosities that are tightly correlated with the width of their lightcurves \citep{Phillips1993ApJ,Phillips+2017book}\footnote{albeit more modern studies find that $\sim30\%$ of SNe Type Ia \\ do not follow this correlation \citep{Taubenberger2017book}.}. This is partly the reason why Type Ia SNe have been traditionally regarded as members of a homogenous population. Conventionally, Type Ia SNe were thought to occur when a carbon oxygen WD approaches the Chandrasekhar mass limit \citep[$\mathrm{M_{ch}}$;][]{Chandra1931ApJ} via accretion. The near-Chandrasekhar mass (near-$\mathrm{M_{ch}}$ henceforth) WD becomes degenerate and thus cannot cool adiabatically in response to continued mass accretion. As a result its central density and temperature increases dramatically, leading to runaway nuclear burning and explosion of the WD \citep{Arnett1969AandSS,Hansen+1969AandSS,Nomoto1982ApJ,Ruiter+2025AnnRev}. 

Early models considered the expanding nuclear flame front to be either strictly subsonic or strictly supersonic; called deflagration \citep{Nomoto+1984ApJ,Iwamoto+1999ApJS} and detonation \citep{Arnett1969AandSS,Hansen+1969AandSS}, respectively. Pure detonation models exhibited very high explosion energies, along with overproduction of iron-group elements (IGEs) \add{and underproduction of intermediate mass elements (IMEs)}, all of which are in conflict with observations \citep{Arnett1971+ApJ}. The deflagration models, on the other hand, suffered from low ejecta kinetic energy and production of lower $^{56}\mathrm{Ni}$ than is required to explain normal Type Ia SNe \citep{Niemeyer+1996ApJ,Niemeyer+1997ApJ}. The ``delayed detonation" class of models were proposed to address this issue. In these models, nuclear burning initiates near the center of a near-$\mathrm{M_{ch}}$ WD as a subsonic flame. It later transitions to develop a supersonic combustion front, or detonation \citep{Ivanova+1974AandSS,Khokhlov+1991AandA_1,Khokhlov+1991AandA_2}. A popular example of this class is the deflagration-to-detonation transition model \citep[DDT; see review by][]{Ropke2017hsn..book}.

Despite the success of these models, the idea of near-$\mathrm{M_{ch}}$ WDs being progenitors of Type Ia SNe were found to suffer from two major issues: they are less abundant in nature compared to their sub-$\mathrm{M_{ch}}$ counterparts \citep{Kepler+2007MNRAS,Torres+2021MNRAS}, and stable growth of WD masses to near-$\mathrm{M_{ch}}$ values requires rather finely-tuned accretion \citep{Nomoto+2007ApJ}. This led to the suggestion that sub-$\mathrm{M_{ch}}$ WDs could also give rise to Type Ia SNe. This is called the double detonation (DD) model, in which a near-surface detonation occurs in the accreted helium shell of a sub-$\mathrm{M_{ch}}$ WD. This detonation sends a shockwave into the core of the WD, causing a second detonation and a subsequent thermonuclear runaway \citep{Livne1990ApJ,Woosley+1994ApJ,Fink2007AandA,Fink+2010A&A,Shen+2014ApJ,Polin+2019ApJ}. Another possible channel for Type Ia SNe from sub-$\mathrm{M_{ch}}$ WDs is the violent merger of two sub-$\mathrm{M_{ch}}$ WDs, followed by carbon ignition near the core of one of the WDs \citep{Pakmor+2010Natur,Pakmor+2012ApJL}.

Both near-$\mathrm{M_{ch}}$ and sub-$\mathrm{M_{ch}}$ WD progenitors are now thought to be needed to explain the range of specific properties of observed Type Ia SNe, such as ejecta mass derived from lightcurve modeling \citep{Scalzo+2014MNRAS,Bora+2024PASP}, Mn/Fe ratio \citep{Seitenzahl+2013AandA,Lach+2020AandA}, nebular spectra-inferred Fe/Ni ratio \citep{Flors+2020MNRAS} and lightcurve-inferred $^{57}\mathrm{Ni}$/$^{56}\mathrm{Ni}$ ratio \citep{Seitenzahl+2009ApJ,Ropke+2012ApJ,Tiwari+2022MNRAS}, with most of the proposed explosion mechanisms likely at play. Perhaps then, it is not surprising that we have moved from the `homogenous family of lightcurves' paradigm to a much more diverse family of Type Ia lightcurves and spectra \citep{Taubenberger2017book,Gal-Yam2017book}, especially with the advent of modern high-cadence surveys such as the Palomar Transient Factory \citep[PTF;][]{Law+2009PASP}, the All-Sky Automated Survey for SuperNovae \citep[ASAS-SN;][]{Kochanek+2017PASP}, the Zwicky Transient Facility (ZTF) Bright Transient Survey \citep{Perley+2020ApJ}, the Young Supernova Experiment \citep[YSE;][]{Aleo+2023ApJS}, the Dark Energy Survey \citep{Abbott+2024ApJ}, and the upcoming Legacy Survey of Space and Time \citep[LSST;][]{Ivezic+2019ApJ}. However, no clear correspondence between observed properties of Type Ia SNe spectra or lightcurves and the proposed thermonuclear explosion mechanisms has been discovered yet \citep{Hillebrandt+2013FrPhy,Ruiter+2025AnnRev}.

Some studies have approached this problem from a different direction, that is, studying remnants of Type Ia SNe to identify the signatures of the mechanism responsible for the explosion \citep{Badenes+2006ApJ,Seitenzahl+2019PhRvL,Ferrand+2021ApJ,Mandal+2025ApJ}. A specific example of such a signature is the presence of a large-scale conical `shadow' in the ejecta of a Type-Ia SNR that exploded via the DD channel in the presence of another WD \citep{Ferrand+2022ApJ}. 

An ideal candidate for this line of investigation is the well-studied supernova remnant (SNR) 0509-67.5, located in the Large Magellanic Cloud (LMC). SNR 0509-67.5 (henceforth 0509) has been observed extensively in radio \citep{Bozzetto+2014MNRAS}, infrared \citep{Seok+2013ApJ}, optical \citep{Helder+2010ApJ,Hovey+2015ApJ}, UV \citep{Ghavamian+2007ApJ} and X-ray bands \citep{Warren+2004ApJ,Kosenko+2008AandA}. The light echo spectrum of 0509 suggests that it belongs to the class of overluminous, slowly fading, 1991T-like SNe \citep{Rest+2008ApJ}. 1D hydrodynamic models with detailed non-equilibrium ionization calculations \citep{Badenes+2008ApJ} also support this conclusion, showing that both the dynamics and the X-ray line flux ratios for 0509 are most consistent with a highly energetic and luminous SN akin to SN 1991T \citep{Phillips+2022ApJ,Phillips+2024ApJS}. Hydrodynamics-based Bayesian analysis of the kinematics of 0509 as revealed by the Hubble Space Telescope \citep{Arunachalam+2022ApJ} suggest a large value of explosion energy ($\sim 1.3\times10^{51}\mathrm{\,erg}$), again favoring a 1991T spectral subtype for this Type Ia SNR. In addition to spectral classification, \cite{Badenes+2008ApJ} also show that 1D delayed detonation (DDT) explosion of near-$\mathrm{M_{ch}}$ WDs are a good match for 0509. \add{Yet another possibility is an SN explosion within a planetary nebula \citep{Soker2025arXiv}.}

On the other hand, \cite{Seitenzahl+2019PhRvL} argue in favor of a sub-$\mathrm{M_{ch}}$ WD explosion. They obtain high-resolution optical observations that reveal highly ionized Fe clumps in 0509. Using detailed time-dependent ionization balance calculations based on analytic SNR models \citep{Truelove+1999ApJS}, they find that the dynamics and strongly ionized state of the Fe clumps make 0509 most consistent with a 1$M_{\odot}$ WD explosion, although a near-$\mathrm{M_{ch}}$ WD explosion with unburnt He in the ejecta cannot be ruled out. They also favor a large explosion energy ($\approx 1.5\times10^{51}\mathrm{erg}$) and note that this energy is available via detonation of a WD with a $0.85M_{\odot}$ carbon-oxygen (CO) core and a $0.15M_{\odot}$ helium (He) shell. Recent optical tomography of 0509 from the deep MUSE(Multi Unit Spectroscopic Explorer) data revealed a double shell [Ca \textsc{xv}] and a single shell [S \textsc{xii}] morphology in the ejecta \citep{Das+2025NatAs}. Their study suggested that this signature can be expected from the double detonation of a CO WD with $\sim 1 M_\odot$ core and a helium shell of $\sim 0.03 M_\odot$.

\begin{figure*}
\centering
\includegraphics[width=0.95\textwidth]{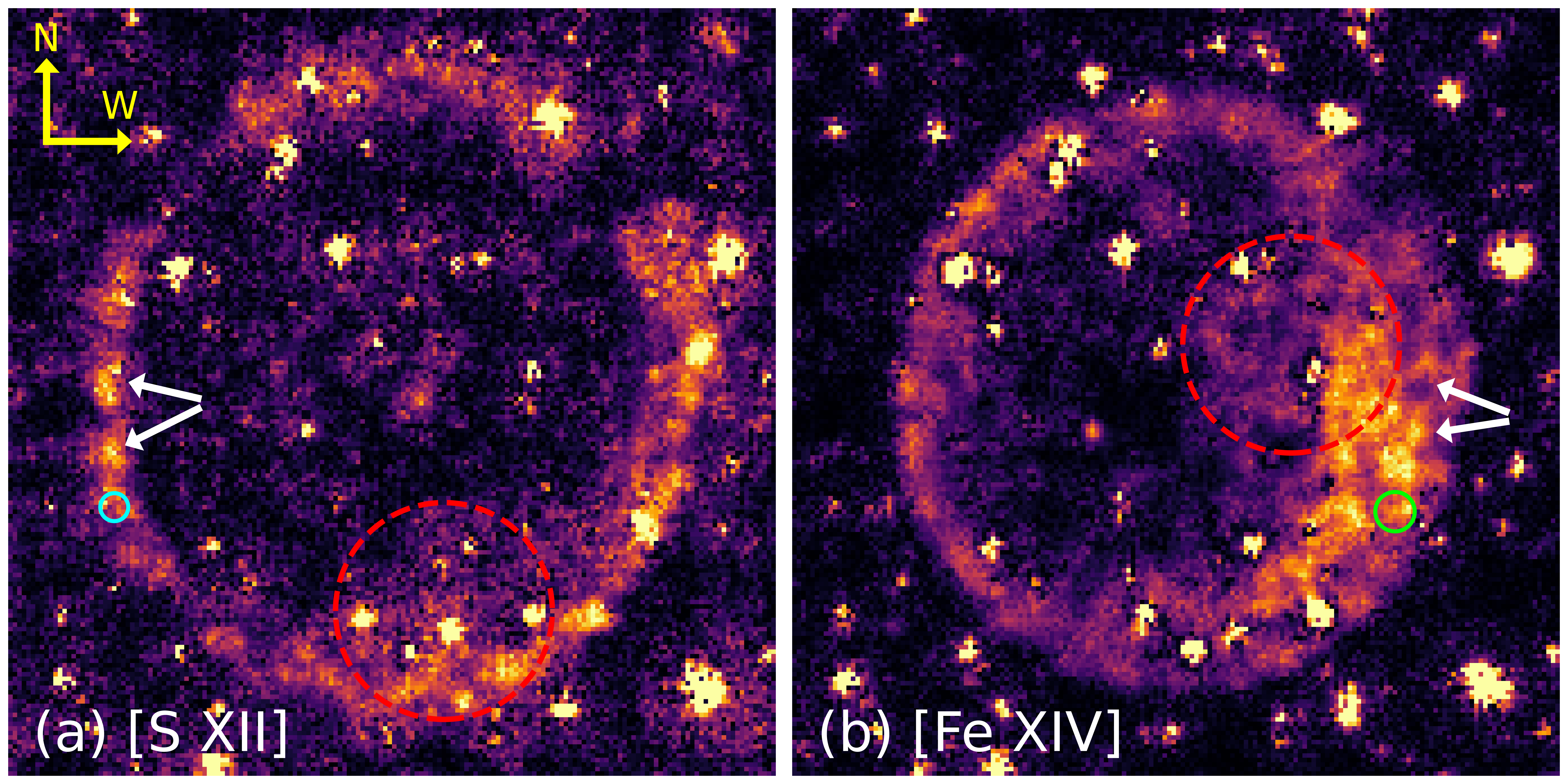}
%\gridline{\fig{data_0509_S.png}{0.32\textwidth}{}\fig{data_0509_Fe.png}{0.32\textwidth}{}}
\caption{[S \textsc{xii}] (\textit{left}) and [Fe \textsc{xiv}] (\textit{right}) emissivity distributions in SNR 0509-67.5. Circles corresponding to spherical harmonics $l\approx30$ (solid cyan) and $l\approx20$ (solid green) are overlaid on the S and Fe images, respectively. A circle of radius corresponding to $l=4$ (dotted red) is also marked on both images. These harmonics correspond to bends in the power spectra of 0509 (shown in Fig.~\ref{fig:ps_0509}) and likely to the small scale substructures in these images. \add{White arows overlaid on the plots point to some examples of these substructures.}}
\label{fig:data_0509}
\end{figure*}

Despite the growing evidence for a sub-$\mathrm{M_{ch}}$ WD progenitor for 0509, sub-$\mathrm{M_{ch}}$ models have generally been considered poor matches to the 1991T-like spectral subtype. \add{Very high peak luminosities ($M_B<-19.5$ mag) and slow decline rates ($\Delta m_{15}[B]= 0.95\pm0.05$) are exhibited by 1991T-like SNe \citep[][]{Phillips+2022ApJ, Phillips+2024ApJS}, implying large $^{56}\mathrm{Ni}$ yields \citep[$1.0-1.4M_{\odot}$; see][]{Filippenko+1992ApJ}. Typically, sub-$\mathrm{M_{ch}}$ explosion models produce less $^{56}\mathrm{Ni}$ \citep[$0.3-0.8M_{\odot}$; e.g.,][]{Sim+2010ApJL} owing to their lower central densities.} Although double detonation models were proposed to explain the peculiar Fe absorption features in pre-maximum phase optical spectra of 1991T-like SNe \citep{Filippenko+1992ApJ}, more recent studies have identified tensions between spectral properties of sub-$\mathrm{M_{ch}}$ models and 1991T-like SNe. In particular, the peak B-band magnitude ($M_B$) and high Si II velocities favor a near-$\mathrm{M_{ch}}$ origin \citep{Polin+2019ApJ,Polin+2021ApJ}, and double detonation models tend to overproduce the [Ca II]/[Fe III] ratio in the optical nebular spectra of luminous (including 1991T-like) SNe \citep{Polin+2021ApJ}.

\add{SNR 0509 provides us an unparalleled opportunity to probe the tension between 1991T-like SNe and DD models, in a way that cannot be pursued with early time SNe. This is due to the fact that SNR 0509 is in a late enough stage of evolution for its ejecta structure to be spatially resolved \citep[$\sim$ 400 yrs;][]{Rest+2005Nature} unlike early time SNe, but not evolved enough for its ejecta to have mixed substantially with gas from the interstellar medium (ISM). This allows us to use a recent result by \cite{Mandal+2025ApJ}, who show that the typical size of small-scale substructures in Type Ia SNRs with sub-$\mathrm{M_{ch}}$ WD progenitors depend strongly on their composition. These substructures originate from turbulence in the SNR, which becomes susceptible to hydrodynamic instabilities (Rayleigh-Taylor Instability and Kelvin-Helmholtz Instability) as it expands against the ambient medium \citep{Chevalier+1978ApJ,Chevalier+1992ApJ}. The size distribution of these substructures bear signatures of the outermost layers of the stellar ejecta \citep{Polin+2022ApJ,Mandal+2023ApJ}.}

\add{This study builds directly on the method proposed by \cite{Mandal+2025ApJ}, focusing on Type Ia SNRs originating from a double detonation explosion. We develop a suite of three-dimensional double detonation SNR models with a range of progenitor masses and obtain substructure size distributions as a function of time for each model. We then analyze optical coronal line maps \citep[forbidden line emission from highly ionized atoms; see][]{Seitenzahl+2019PhRvL} of SNR 0509 (Fig.~\ref{fig:data_0509}) and compare the resulting substructure morphologies with those of our numerical models. This provides an application of the method of \cite{Mandal+2025ApJ} to a well-observed remnant. It yields an independent morphological test of the explosion mechanism of 0509, complementary to the analysis of element distributions by \cite{Das+2025NatAs}. It also allows us to place constraints on the core and shell masses of the progenitor WD.}

This work is organized as follows: The observational data used in this work are described in Section \ref{sec:observations}. Details of the SNR models and the techniques used to analyze the models and the data are discussed in Section \ref{sec:numerical}. The results are presented in Section \ref{sec:results}, and  a summary of the results is presented in Section \ref{sec:discussion}.

\section{Observational data}     \label{sec:observations}
A deep observational campaign targeting the supernova remnant SNR 0509 was carried out using the Multi Unit Spectroscopic Explorer \citep[MUSE;][]{bacon2010muse}, an optical integral field spectrograph mounted on Unit Telescope 4 (UT4) of the European Southern Observatory’s (ESO) Very Large Telescope at Cerro Paranal. This effort was part of ESO program 0104.D-0104(A), led by P.I. Seitenzahl. Observations were conducted in service mode using the Wide Field Mode with Adaptive Optics (WFM-AO) configuration, spread over 25 nights across a 24-month period  \citep[see Methods in][]{Das+2025NatAs}. A total of 40 exposures were obtained: 39 of these were approximately 2700 seconds in duration, while one shorter observation lasting 93.92 seconds was discarded due to poor data quality. The resulting dataset comprises a total exposure time of roughly 105,300 seconds (equivalent to 29 hours and 15 minutes). Each MUSE data cube covers the optical wavelength range from $4690\,\mathrm{\AA}$ to $9340\,\mathrm{\AA}$ at a spectral resolution of R$\sim$3000. To minimize contamination from the laser guide stars of the 4LGSF system, a notch filter is employed in WFM-AO mode to block the spectral region from $5804\,\mathrm{\AA}$ to $5965\,\mathrm{\AA}$ \citep{vogt2018raman,Vogt2017}. However, this filter does not completely eliminate the effects of Raman-scattered photons originating from the laser light. These residual artifacts are subsequently identified and removed during data processing using the MUSE data reduction pipeline \citep{Weilbacher2020}.

Data reduction was performed using ESOReflex \citep{freudling2013} version 2.11.5 and the MUSE data reduction pipeline version 2.8.9 \citep{Weilbacher2020}, both run with default settings. The process included the removal of known sky lines and was carried out on Tycho, a high-memory Linux workstation at the University of New South Wales in Canberra, specifically configured for handling MUSE data. Since no dedicated sky observations were available, background subtraction was performed using the pipeline's alternative method: computing a sky model from a specified portion of the field of view, controlled by the parameter SkyFr\_2. All individual data cubes were stored in a single directory, used as the pipeline’s input (with no additional calibration files). The final mosaic was produced by re-running the pipeline with the 39 reduced MUSE pixel tables as input, combining them into a single data cube. This mosaic covers a $1\mathrm{\arcmin} \times 1\mathrm{\arcmin}$ field of view, with a spatial sampling of $0.2\mathrm{\arcsec} \times 0.2\mathrm{\arcsec}$ per spaxel, and was used for all subsequent analysis.

\add{For this work, we choose the [Fe \textsc{xiv}] $\lambda$5303 \AA\, on account of being the brightest of all the coronal lines and [S \textsc{xii}] $\lambda$7611 \AA\, for being the brightest non-iron coronal line. \cite{Das+2025NatAs} provide a representative spectrum for 0509 (extracted from a region on its western side; see their Fig.~1). However, the overall surface brightness of the [Fe \textsc{xiv}] $\lambda$5303 \AA\, and [S \textsc{xii}] $\lambda$7611 \AA\, lines is much lower than that of H$\alpha$, which is dominated by strongly limb-brightened Balmer filaments along the outer rim of the remnant. Hence, conventional background subtraction was found to be insufficient. The final mosaic data cube was thus processed with an additional background subtraction and galactic de-reddening (\citealp[see details]{Das+2025NatAs, Ghavamian2017}).}

[Fe \textsc{xiv}] $\lambda$5303 \AA\, and [S \textsc{xii}] $\lambda$7611 \AA\, emission are visualized by integrating the spectra over specific wavelength intervals ($\lambda_1$ and $\lambda_2$) corresponding to the emission features of each forbidden line. To remove the underlying stellar continuum, adjacent spectral regions on either side of the emission line ($\Delta \lambda_{*1}$ and $\Delta \lambda_{*2}$) are also integrated and subtracted from the signal. The chosen wavelength ranges for both the emission features and the surrounding continua are listed in Table 1.

\begin{deluxetable*}{cccccc}
\tablecaption{Spectral windows used to remove stellar continuum and isolate emission from ionized species.} 
\label{table:observation}
\tablehead{
\colhead{Coronal} & \colhead{$\lambda_1 (\textrm{\AA})$} & \colhead{$\lambda_2 (\textrm{\AA})$} & \colhead{$\Delta \lambda_{*1} (\textrm{\AA})$} & \colhead{$\Delta \lambda_{*2} (\textrm{\AA})$} & \colhead{Peak Surface Brightness}\\[-2ex] 
\colhead{Emission Lines} & \colhead{} & \colhead{} & \colhead{} & \colhead{} & \colhead{($\mathrm{erg s^{-1}cm^{-2}arcsec^{-2}}$)} \\[-2ex]
}
\startdata
    [Fe \textsc{xiv}]  & 5218 & 5364  & 5150 - 5190 & 5485 - 5528  & $6.77\times 10^{-17}$ \\ \hline
     [S \textsc{xii}]   & 7602 & 7705  & 7393 - 7484 & 7779 - 7809  & $2.67\times 10^{-17}$ \\
\enddata

\end{deluxetable*}

\section{Numerical method}  \label{sec:numerical}

\subsection{Initial conditions} \label{subsec:initial}

We use angle-averaged (1D) versions of DD models described in \cite{Gronow+2021AandA}, hosted on the HESMA project \citep{Kromer+2017MmSAI} website\footnote{\hyperlink{https://hesma.h-its.org}{https://hesma.h-its.org}}. These models span a range of CO core masses and He shell masses. The details of these models are listed in Table~\ref{table:sn_models}. \add{The angle-averaged profiles reproduce the radial distribution of mass and composition in the ejecta but do not retain the asymmetries intrinsic in the original 3D explosion models. Studies that follow fully three-dimensional supernova models into the remnant phase have shown that such explosion-imprinted anisotropies primarily influence the global morphology of the remnant, corresponding to the lowest spherical harmonic modes ($l \lesssim 10$), whereas the structures at intermediate and smaller angular scales arise mainly from hydrodynamic instabilities that develop as the remnant evolves \citep{Ferrand+2019ApJ,Ferrand+2021ApJ}.}

\add{Additional tests presented by \cite{Mandal+2023ApJ} examined how perturbations introduced in the ejecta (or the ambient medium) propagate during remnant evolution. These calculations show that initially imposed perturbations do not typically produce identifiable features in the power spectrum for long. Instead, within a few decades the structure of the remnant becomes governed by Rayleigh–Taylor instability growth at the contact discontinuity. Because the present study focuses on the statistical properties of structures that emerge during the SNR phase (several centuries after the explosion), we do not expect the omission of explosion-imprinted perturbations in the initial conditions to significantly influence the measured power spectra.}

The ejecta density profile and abundance profiles for S and Fe from the SN models are mapped onto a 3D domain and allowed to expand to the SNR phase (described in the next Section). The ejecta are assumed to be cold ($P = 10^{-6}\rho$) and expanding homologously ($v=r/t$). The ambient medium is set to have uniform density. \add{Using spherically symmetric initial conditions also allows us to model only one octant of the remnant \citep[as in][]{Mandal+2023ApJ} without introducing additional nonphysical symmetries, enabling higher effective resolution than would be practical for a full-sphere simulation at comparable computational cost.}

%$n_0 = 0.2\mathrm{cm}^{-3}$

%\add{We solve for an octant of the spherical SNR following \cite{Mandal+2023ApJ}, who show that the turbulent substructures formed in this case are identical to those in a fully spherical model. Moreover, solving for  effectively doubles the resolution compared to solving for a full sphere}

%Solving for an octant effectively doubles the resolution compared to solving for a full sphere. This is a unique advantage of using angle-averaged initial conditions over 3D SN models. In that case, selecting some octant of the model and evolving it to the SNR phase may give rise to unphysical angular modes due to the imposed symmetry. These modes may conflict with the turbulent peak in the power spectrum.

%\add{These angle-averaged initial conditions are free of anisotropies inherent to the ejecta. It has been shown that such anisotropies mostly influence the large-scale structure of the SNR, but are not significant contributors to the size distribution of its turbulent substructures \citep{Ferrand+2019ApJ,Mandal+2023ApJ}. Hence, we do not expect the absence of such ejecta-intrinsic anisotropies to affect our studies. In addition, spherical symmetry of the initial conditions allows us to model only an octant of the spherical SNR \citep[as in][]{Mandal+2023ApJ}, without imposing additional symmetries on the system. This effectively doubles the resolution compared to solving for a full sphere.}

\begin{deluxetable}{cccc}
\tablecaption{Summary of the DD SN models used in this work.} 
\label{table:sn_models}
\tablehead{
\colhead{Name} & \colhead{Core mass} & \colhead{Shell mass} & \colhead{Explosion energy} \\[-2ex]
\colhead{}     & \colhead{(in $M_{\odot}$)} & \colhead{(in $M_{\odot}$)} & \colhead{(in $10^{51}$ erg)}
%\\  \colhead{(1)} & \colhead{(2)} & \colhead{(3)}  
}
\startdata
M0803    & 0.8 & 0.03 & 0.75 \\
M0805    & 0.8 & 0.05 & 0.84 \\
M0903    & 0.9 & 0.03 & 0.97 \\
M0905    & 0.9 & 0.05 & 1.05 \\
M0910    & 0.9 & 0.10 & 1.50 \\
M1003    & 1.0 & 0.03 & 1.44 \\
M1005    & 1.0 & 0.05 & 1.42 \\
M1010    & 1.0 & 0.10 & 1.48 \\
\enddata
%\tablenotetext{a}{ mass of the ejecta; excludes mass of the bound remnant wherever present}{b}{ only model in the paper.}
\end{deluxetable}

\subsection{Remnant evolution and analysis} \label{subsec:remnant}

The 1D models are mapped onto a 3D Cartesian grid in \sprout, a second-order ideal hydrodynamics code \citep{Mandal+2023_sprout} designed to studying expanding outflows. \sprout\; utilizes the moving mesh paradigm \citep{Springel2010MNRAS,Duffell+2011ApJS} to expand its computational domain with time, following the evolution of the SNR. The expanding grid provides a high dynamic range in resolution. This is necessary to model the SNR's expansion, which starts at $100s$ and continues till $600\,\mathrm{yrs}$. The end time is chosen keeping in mind the age estimate of SNR 0509-67.5, which is roughly in the range $300-500\,\mathrm{yrs}$ \citep{Rest+2005Nature,Hovey+2015ApJ}.

It is convenient to describe the time evolution of the SNR models in terms of a dimensionless quantity $t_D$, called the dynamical age of the remnant. This is essentially the age of the remnant ($t$) divided by the characteristic time or Sedov time ($T_c$), which is the time taken by the forward shock to sweep up mass comparable to the ejecta mass itself. The characteristic time is a function of the ejecta mass $M$, explosion energy $E$, and ambient medium (AM) density \citep{Truelove+1999ApJS,Warren+2013MNRAS,Mandal+2024ApJ}. Using this, we can define the dynamical age $t_{D}$ as follows:

\begin{figure*}
\centering
\gridline{ \fig{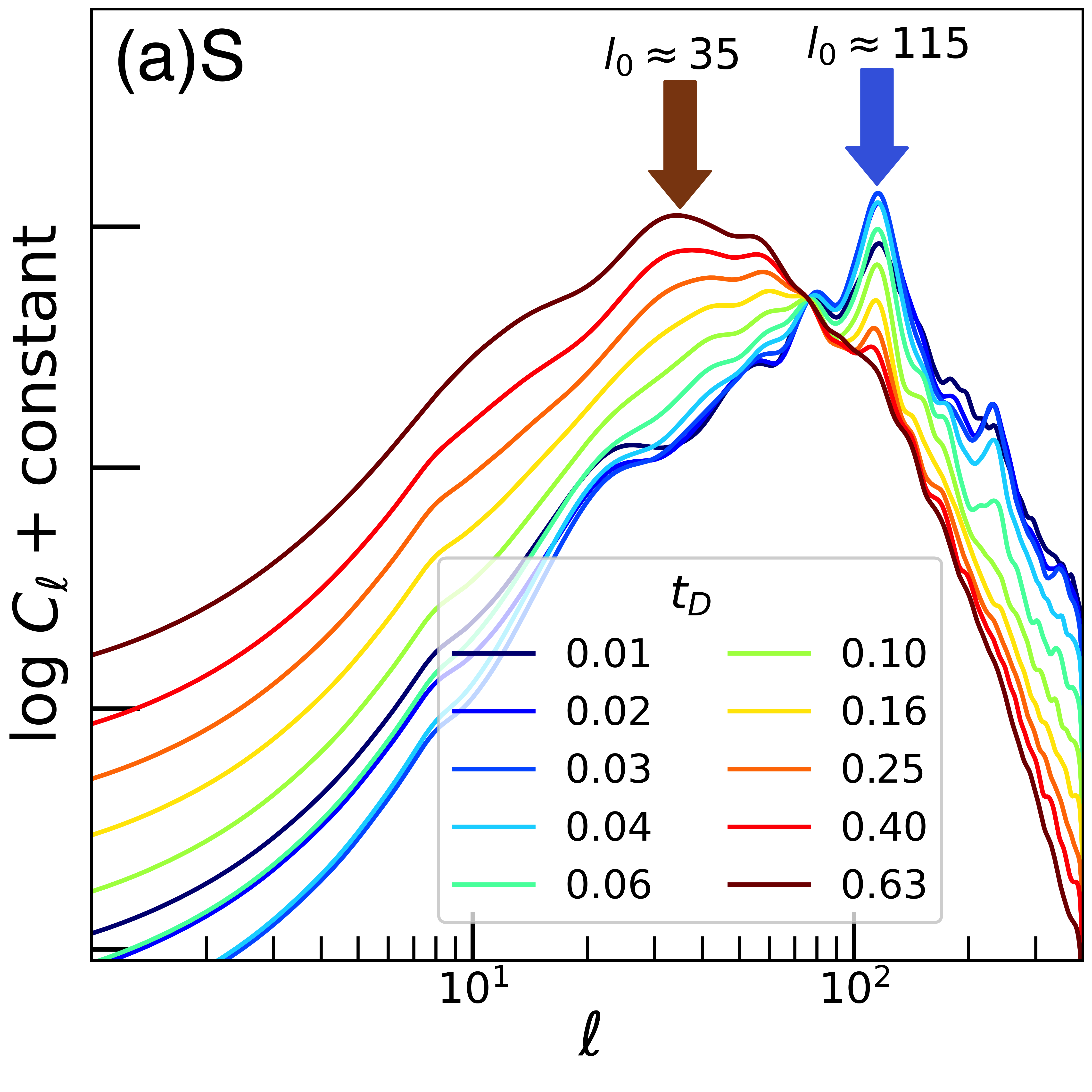}{0.499\textwidth}{}
          \fig{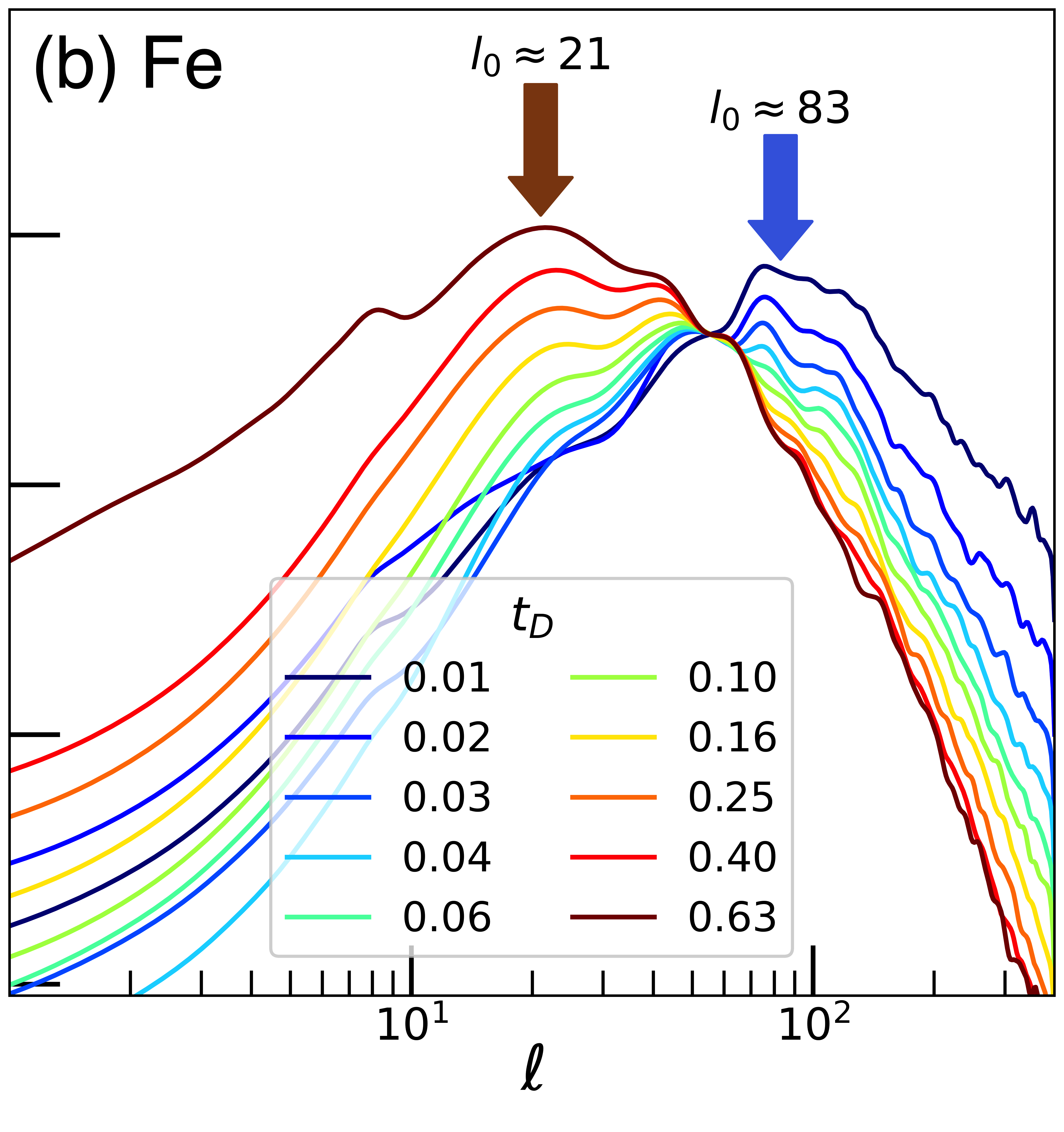}{0.469\textwidth}{}}
\vspace{-10mm}
\caption{Power spectrum of S (\textit{left}) and Fe (\textit{right}) distributions in the ejecta of the M1003 model, evaluated at 10 logarithmically spaced time instants in the range $t_D=0.01-0.63$. All power spectra have a broken power-law shape. The wavenumber where the power spectrum changes slope or peaks ($l_0$) corresponds to the lengthscale where most of the power in the turbulent substructures reside. With time, $l_0$ shifts to smaller values, indicating growth of the \add{turbulent} substructures. For DD SNRs, $l_0$ of S distribution (\los henceforth) consistently remains larger than the $l_0$ for Fe distribution (\lofe henceforth). For example, at $t_D=0.01$, \los$\approx115$, while \lofe$\approx83$. \add{The overall normalization for each power spectrum were chosen arbitrarily, to aid visual inspection and easy identification of $l_0$.}}
\label{fig:ps_model}
\end{figure*}

\begin{equation}
\label{eq:dynamical_age}
        t_{D} \equiv \left(\frac{t}{628\mathrm{yrs}}\right)\left(\frac{M}{M_{ch}}\right)^{5/6} E_{51}^{-1/2}n_0^{-1/3},
\end{equation}

where $M_{ch}=1.4M_{\odot}$ is the Chandrasekhar mass, $E_{51}=E/(10^{51}\mathrm{ergs}$), and $n_0$ is the AM density in units of $\mathrm{amu\,cm^{-3}}$. Note that this definition is not unique. For example, \cite{Warren+2013MNRAS} choose a different characteristic age $T'$, which is related to our choice as $T' = 0.43T_c$.

%, and $n_0 =\rho_0/(2.34\times10^{-24}\mathrm{g}$) is the number density of the circumstellar medium, assuming a 10:1 H:He ratio

We compute power spectra of these 3D models at several instants of time, using the method outlined in \cite{Polin+2022ApJ} and \cite{Mandal+2023ApJ}. \add{Since the computations evolve only one octant of the remnant, the data for each model snapshot is reflected across the coordinate planes to reconstruct a full spherical remnant. This imposed symmetry affects only the lowest multipoles, which are not used in the present analysis.} Spherical surface distributions of the radially integrated density distributions of \add{S} (and Fe) are then calculated for each reconstructed snapshot as follows:

\begin{equation}
        \left<\rho_{S}\right>(\theta,\phi) = \frac{\int X_{S} P \rho dr}{\int X_{S} P dr}, 
\label{eq:surface_map}
\end{equation}

where $X_{S}$ is the passive scalar representing the mass fraction of \add{S}. These spherical surface distributions are expanded in terms of spherical harmonics using the SHTOOLS package \citep{SHTOOLS}. This allows us to derive a power spectrum $C_l$ in terms of the spherical harmonic $l$ \citep[see Eqns.~11 and 12 of][]{Mandal+2024ApJ}. \add{Weighting by the product of pressure and the relevant passive scalar (tracing S or Fe ejecta) ensures that only the shocked, turbulent ejecta contributes to this calculation. Within the region between the forward and reverse shocks, this product varies by at most a factor of $\lesssim1.5$, so the weighting does not strongly bias the calculation toward localized high-pressure regions. To verify that our results are not sensitive to the precise weighting prescription, we repeat the analysis using a constant weight for all cells above a pressure threshold. Power spectra derived from the spherical surface maps for each case are found to be nearly identical.}

\add{Our models utilize discretization errors associated with the Cartesian grid to provide seed perturbations to the contact discontinuity, which rapidly becomes unstable and exhibits growth of turbulent substructures. As found by \cite{Polin+2022ApJ} and \cite{Mandal+2023ApJ}, the nature of seed perturbations (whether inherent to the grid or imposed externally) does not appreciably affect the growth of turbulence in SNRs. \cite{Mandal+2023ApJ} further show (in their Appendix) that grid-scale features typically appear in the power spectrum of SNR models at very large spherical harmonics ($l\sim1000$ for a grid with 512 cells per dimension) and do not interfere with the lengthscales that correspond to the typical size of turbulent substructures ($l\lesssim100$). Following \cite{Mandal+2023ApJ}, we adopt a resolution of $512^3$ cells for our computational grid to ensure minimal influence of grid-scale noise on the power spectra.}

%Within the region between the forward and reverse shocks, this product varies by at most a factor of ≲2.5, so the weighting does not strongly bias the calculation toward localized high-pressure regions. To verify that our results are not sensitive to the precise weighting prescription, we repeated the analysis using a constant weight for all cells above a pressure threshold; the resulting power spectra are nearly identical. We therefore conclude that the inferred spectral properties are robust to the choice of weighting scheme.

\subsection{Image analysis} \label{subsec:image_anly}

The substructure size distributions in the images of 0509 are measured using the $\Delta$-variance algorithm \citep{Arevalo+2012MNRAS}, as adapted by \cite{Mandal+2024ApJ}. This technique computes a low resolution power spectrum of a two-dimensional image, as is the case here. It is to be noted that this power spectrum is not necessarily equivalent to those of 3D models we describe in Section~\ref{subsec:remnant}. However, \cite{Mandal+2024ApJ} showed that power spectra of their 3D SNR models are equivalent to the $\Delta$-variance-based power spectra of synthetic images corresponding to the same models, for spherical harmonics $l\gtrsim10$.
\add{This equivalence provides the foundation for our analysis. We thus interpret the $\Delta$-variance–derived power spectra of 0509 as representative of the underlying 3D ejecta distribution and directly compare them to the power spectra of our 3D models.}
%Therefore, in this work we treat the power spectra of 0509 images as those that one would obtain upon analysis of the full 3D distribution of the ejecta in 0509. In other words, the $\Delta$-variance-based power spectra of 0509 images are directly compared to those of our 3D models. 

Since $\Delta$-variance computes power as a function of lengthscale $\sigma$ (in pixel units), it has to be converted to the angular harmonic $l$ as shown by \cite{Mandal+2024ApJ}:

\vspace{-2mm}

\begin{equation}
\label{eq:sigma_to_l}
    l\approx 1.6R_{\mathrm{SNR}}/\sigma,
\end{equation}

where $R_{\mathrm{SNR}}$ is the radius of the image of the SNR, also in pixel units. We apply masks on the \add{field stars} to ensure they are not contaminating our analysis.

%%%**********

\section{Results}   \label{sec:results}

\subsection{Power spectral analysis of numerical models}
\label{subsec:ps_models}

Power spectra of hydrodynamic SNR models without large scale asymmetries are expected to have a broken power-law form \citep{Warren+2013MNRAS,Polin+2022ApJ,Mandal+2023ApJ}\footnote{There can be deviations from this shape if large scale\\ asymmetries are present, see \cite{Ferrand+2021ApJ} for examples.}:

\begin{equation}
\label{eq:BPL}
    C_l \propto \frac{1}{(l/l_0)^{-n_1}+(l/l_0)^{n_2}},
\end{equation}

where $l_0$ is the wavenumber at which the power spectrum peaks. The power spectra can be approximated at $l\lesssim l_0$ as $C_l\propto l^{n_1}$, and at $l\gtrsim l_0$ as $C_l\propto l^{-n_2}$, with $n_1$ and $n_2$ being positive. The peak wavenumber $l_0$ corresponds to the typical size of the turbulent eddies formed in the ejecta-AM interaction of the SNR. 

Our double detonation SNR models show a similar behavior. In Fig.~\ref{fig:ps_model}, we plot the power spectra of S (left panel) and Fe (right panel) distributions in the M1003 model at various instants of time. The power spectra of S and Fe represent those of IMEs and IGEs, respectively. \add{The overall normalization of each power spectrum is arbitrary and were chosen to aid visual inspection.} Power spectra at early times are plotted in blue, and the colors transition to red at late times. For both panels, $l_0$ decreases with time, that is, the peak shifts to the left. This behavior can be understood by noting that $l_0$ is roughly proportional to the effective power-law slope of the ejecta encountered by the reverse shock \citep{Polin+2022ApJ,Mandal+2023ApJ}. This result suggests that the typical size of turbulent eddies in SNRs is of the order of the density scale height at the reverse shock. Since the ejecta density falls off steeply at large radii but is relatively shallow in the inner regions \citep[for an example, see Fig.~2 of][]{Collins+2025MNRAS}, the reverse shock encounters shallower density profiles as it moves deeper into the ejecta. Thus the value of the ejecta power-law slope encountered by the reverse shock \add{decreases} with age.

\add{The value of $l_0$ thus allows us to constrain the dynamical age of an SNR, as will be further demonstrated in section~\ref{subsec:model_vs_0509}. It is also possible to measure $l_0$ from the density distribution of the ejecta (not differentiating between elements) and estimate the dynamical age of the SNR \citep{Mandal+2024ApJ,Mandal+2025ApJ}.} Moreover, Fig.~\ref{fig:ps_model} shows that the value of $l_0$ for the power spectrum of IMEs (\los\, henceforth) is seen to be greater than the value of $l_0$ for the power spectrum of IGEs (\lofe\, henceforth) at every instant of time. This is an exclusive signature of SNRs with sub-$\mathrm{M_{ch}}$ WD progenitors \citep{Mandal+2025ApJ}. This property stems from the fact that near-$\mathrm{M_{ch}}$ Type Ia SN progenitors generate a lot of buoyant ashes from their initial deflagration phase that cause significant mixing in the ejecta. The sub-$\mathrm{M_{ch}}$ WD progenitors are expected to experience detonations (or supersonic nuclear flame fronts) only, that do not cause this amount of mixing and therefore retain stratification in their ejecta. This causes the IMEs and IGEs to participate at different times in the ejecta-AM interaction and therefore develop substructures with different typical sizes \citep[see Section 4.4 of][]{Mandal+2025ApJ}. This holds true for all our DD SNR models, as will be seen in section~\ref{subsec:model_vs_0509}.

%The angular power spectrum from each of our models takes on a broken power-law shape starting from the very early stages of evolution, as has been demonstrated before \citep{Polin+2022ApJ,Mandal+2023ApJ}. The break in the shape of the power spectrum (or transition between the two power-laws) occurs at a wavenumber $l_0$, which is also the peak of the power spectrum in the absence of other large-scale seed anisotropies. The wavenumber $l_0$ therefore corresponds to a lengthscale where most of the turbulent power in the SNR accumulates. Figure \ref{fig:ps_example} shows example power spectra for S and Fe distributions from our model M1010 at different instants of time (with blue and red representing early and late times, respectively). For small values of the wavenumber $l$ (corresponding to large length scales), the shape of the power spectrum is can be described as $C_l \propto l^3$, where $C_l$ is the power at wavenumber $l$. This is consistent with analytically obtained linear growth rate of perturbation modes in expanding blastwaves \citep{Chevalier+1992ApJ}. In other words, large scale eddies grow approximately linearly as they evolve.

\subsection{Substructure analysis of SNR 0509-67.5}
\label{subsec:ps_0509}

%The left panel of Fig.~\ref{fig:ps_0509} shows the power spectra obtained from the [Fe \textsc{xiv}] and [S \textsc{xii}] images of 0509. Both of them have strong bumps or peaks at $l=4$. The presence of large power at such a small wavenumber evidently comes from the large scale asymmetry visible in 0509. In addition, the Fe power spectrum shows a break or change in slope at $l\sim20$, while the S power spectrum shows a break at $l\sim30$. 

\begin{figure*}
\centering
\includegraphics[width=0.49\textwidth]{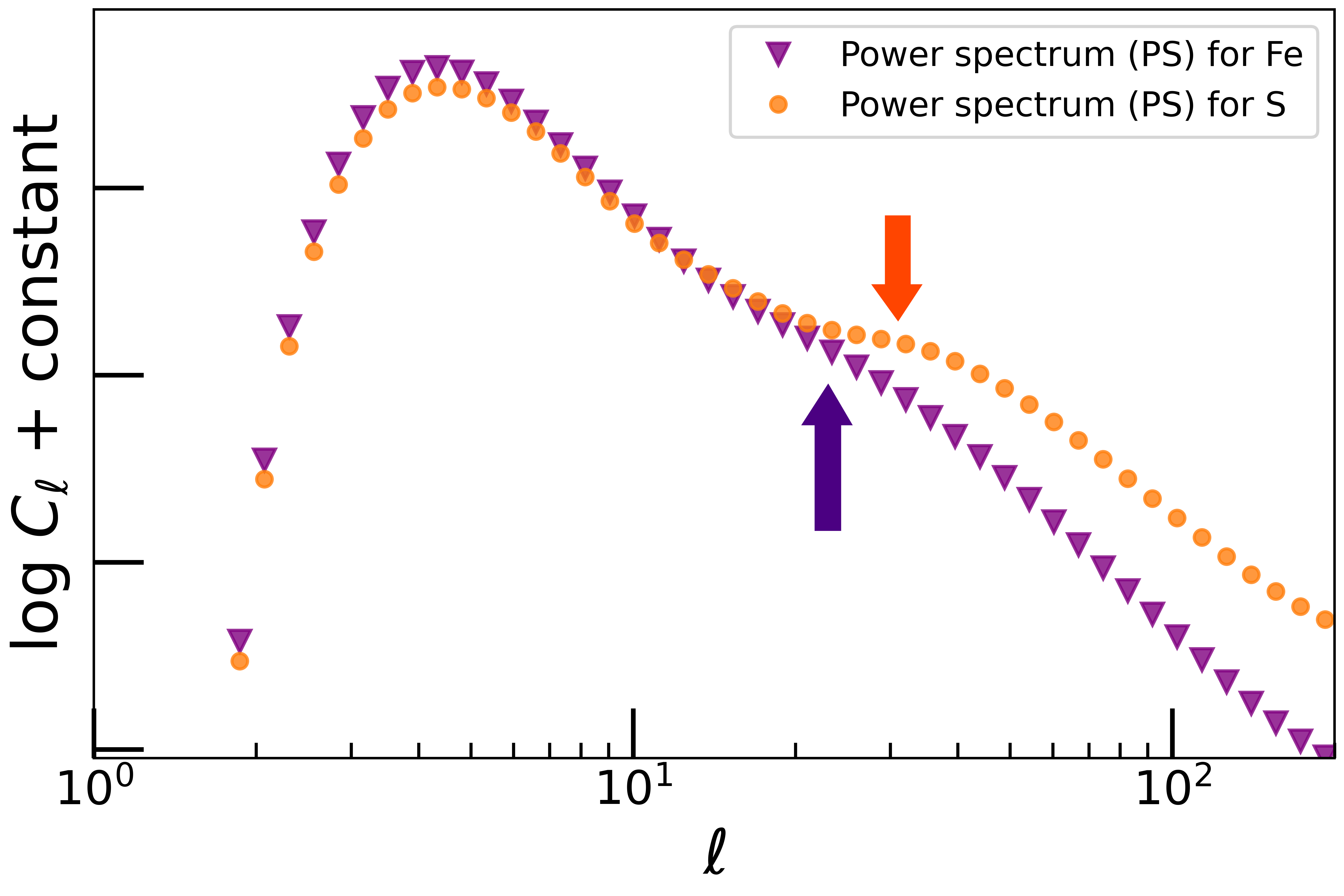}
\includegraphics[width=0.49\textwidth]{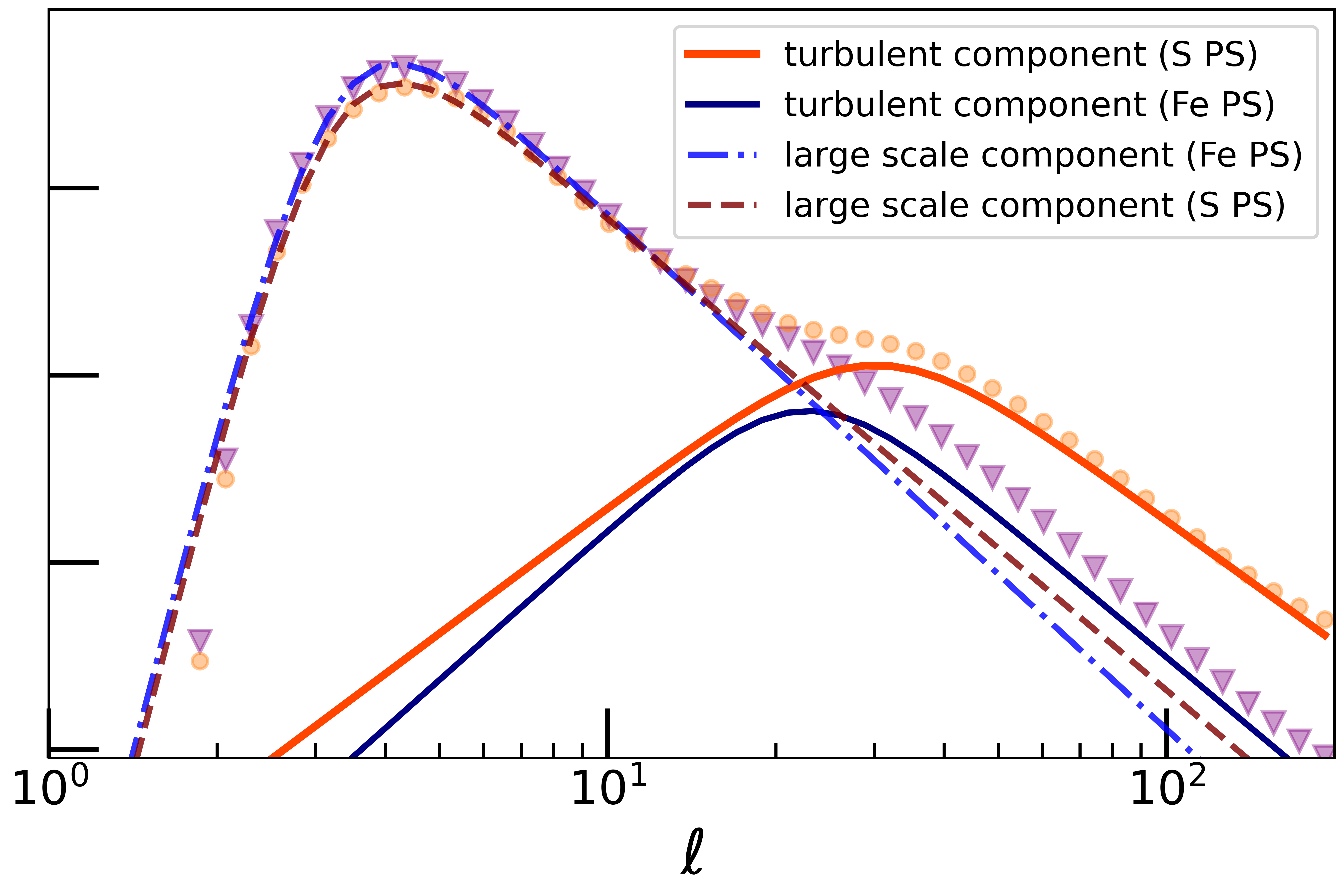}
\caption{\textit{Left}. Power spectra of S and Fe distributions in SNR 0509-67.5. Both power spectra show a strong bump at $l\approx4$, which is anticipated given the presence of the prominent large scale asymmetric features in 0509. Both curves also show a bend or change in slope (at $l\approx30$ and $l\approx20$ for S and Fe respectively, marked by arrows in the plot). These spectral breaks likely correspond to small-scale substructures seen in 0509 (see Fig.~\ref{fig:data_0509}). \textit{Right}. Both power spectra are fit to a two-component model; one corresponding to low-$l$ anisotropies in the ejecta, and the other signifying turbulent activity in the SNR at higher values of $l$ (small-scale substructures).}
\label{fig:ps_0509}
\end{figure*}

\add{The left panel of Fig.~\ref{fig:ps_0509} shows the power spectra obtained from the [Fe \textsc{xiv}] and [S \textsc{xii}] images of 0509. Both spectra exhibit a pronounced peak at $l=4$. A peak at low wavenumbers indicates substantial large-scale asymmetry in the SNR, consistent with the morphology seen in both our optical coronal line maps (Fig.~\ref{fig:data_0509}) and X-ray images of 0509 \citep[see panel 1 of Fig. 1 in][]{Lopez+2011ApJ}. Using Eqn.~\ref{eq:sigma_to_l}, we illustrate the characteristic scale of the $l=4$ peak by overlaying the Fe and S maps in Fig.~\ref{fig:data_0509} with a dotted red circle of the corresponding diameter, highlighting structures on comparable scales.}

%The presence of large power at such a small wavenumber evidently comes from the large scale asymmetry visible in 0509. In addition, the Fe power spectrum shows a break or change in slope at $l\sim20$, while the S power spectrum shows a break at $l\sim30$. Following \cite{Mandal+2024ApJ,Mandal+2025ApJ}, we interpret these breaks as the turbulent peak $l_0$ for the Fe and S distributions, respectively. Either power spectrum for 0509 may thus be viewed as a broken power-law corresponding to turbulent activity, plus a large bump at $l\approx4$ accounting for large-scale asymmetries inherent to the ejecta. As a sanity check for this interpretation, we overlay the S and Fe distribution images in Fig.~\ref{fig:data_0509} with circles of radii corresponding to $l = 4,20,\mathrm{\,and\,}30$, calculated using Eqn.~\ref{eq:sigma_to_l}. Visual inspection shows that the bump at $l=4$ likely corresponds to prominent large scale structures (likely locations marked with red circles in Fig.~\ref{fig:data_0509}) in 0509. Moreover, the lengthscales corresponding to $l=20$ and $l=30$ appear to be similar to the small-scale substructures in 0509 (see, e.g., the alternately bright and dark circular patches in northwestern part of the [Fe XIV] emission in Fig.~\ref{fig:data_0509}b).

\add{In addition to the large-scale asymmetry traced by the $l=4$ peak, both power spectra exhibit clear breaks in slope at higher wavenumbers (marked with arrows in the left panel of Fig.~\ref{fig:ps_0509}. The Fe power spectrum shows a break near $l\approx20$, while the S power spectrum displays a similar feature near $l\approx30$. These wavenumbers correspond to smaller characteristic physical scales, indicative of compact substructures in 0509. Notably, a qualitatively similar spectral morphology has been reported for Tycho’s SNR \citep{Mandal+2024ApJ,Mandal+2025ApJ}, including the low-$l$ peak and the higher-$l$ break. Motivated by this correspondence, we associate these physical scales with turbulent substructures in 0509. To illustrate this possible correspondence, we overlay circles of the corresponding diameters on the 0509 images (Fig.~\ref{fig:data_0509}). These broadly match the sizes of the observed small-scale substructures in 0509 (see, e.g., the alternately bright and dark circular patches in northwestern part of the Fe image in Fig.~\ref{fig:data_0509}).}

%Based on the arguments above, we adopt the view that the breaks in the power spectra of 0509 correspond to the turbulent peaks in the power spectra of our SNR models. To identify these break wavenumbers accurately, we fit the power spectra of S and Fe distributions in 0509 to the sum of two broken power laws (both components are as in Eqn.~\ref{eq:BPL}), one corresponding to turbulent activity and the other to large scale intrinsic asymmetry in the ejecta. For the power spectrum of S distribution, the fit provides $l_0 (= l_0^{\mathrm{S}}) = 31\pm2$ and $n_2 = 2.16\pm0.05$. For the power spectrum of Fe distribution, the fit provides $l_0 = (l_0^{\mathrm{Fe}}) = 22\pm1$ and $n_2 = 2.50\pm0.04$. Considering the radius of 0509 as observed in the sky is $\approx15''$ \citep{Arunachalam+2022ApJ}, we find the typical size of S and Fe dominated substructures seen in Fig.~\ref{fig:data_0509} to be $\approx0.8''$ and $\approx1.1''$ respectively (using Eqn.~\ref{eq:sigma_to_l} and the values of \los and \lofe). 

\add{Motivated by the arguments above, we interpret the large-$l$ power spectral breaks in 0509 as the turbulent peaks seen in the power spectra of our SNR models. According to this interpretation, each observed power spectrum may be described as the superposition of two components: a large-scale feature near $l\approx4$ arising from large-scale intrinsic ejecta asymmetries, and a broken power-law component (as in Fig.~\ref{fig:ps_model} or Eqn.~\ref{eq:BPL}) associated with turbulent activity. To quantify the break wavenumbers, we therefore fit the Fe and S power spectra of 0509 with the sum of two broken power laws (each of the form given in Eq.~\ref{eq:BPL}), representing the asymmetry and turbulent contributions, respectively. For the power spectrum of S distribution, the fit provides $l_0 (= l_0^{\mathrm{S}}) = 31\pm2$ and $n_2 = 2.16\pm0.05$. For the power spectrum of Fe distribution, the fit provides $l_0 = (l_0^{\mathrm{Fe}}) = 22\pm1$ and $n_2 = 2.50\pm0.04$, with the quoted uncertainties corresponding to $1\sigma$ confidence intervals for either case. Considering the radius of 0509 as observed in the sky is $\approx15''$ \citep{Arunachalam+2022ApJ}, we find the typical size of S and Fe dominated substructures seen in Fig.~\ref{fig:data_0509} to be $\approx0.8''$ and $\approx1.1''$ respectively (using Eqn.~\ref{eq:sigma_to_l} and the values of \los and \lofe).}

We find $l_0^{\mathrm{S}}/l_0^{\mathrm{Fe}}=1.42\pm0.20$, that is, S dominated substructures in 0509 have a larger value of $l_0$ and thus a smaller typical size compared to their Fe dominated counterparts. This is exactly the signature of Type Ia SNRs with sub-$\mathrm{M_{ch}}$ progenitors found by \cite{Mandal+2025ApJ}, \add{as mentioned in Section~\ref{subsec:ps_models}. In addition, both the S and Fe power spectra at small scales obey a power law $C_l\propto l^{-n_2}$, with $n_2>2$. This is much steeper compared to the power spectrum expected for a Kolmogorov-like turbulent cascade \citep[$C_l\propto l^{-5/3}$;][]{Kolmogorov1941DoSSR}. Hence, 0509 also exhibits an absence of turbulent cascade at small scales that is typically seen in numerical SNR models \citep{Polin+2022ApJ,Mandal+2024ApJ}. }

%As mentioned in Section~\ref{sec:intro}, \cite{Mandal+2025ApJ} show that such values of the power spectra peak wavenumber ratio (of IMEs to IGEs) favor a sub-$\mathrm{M_{ch}}$ WD progenitor, while an SNR with a near-$\mathrm{M_{ch}}$ WD progenitor would have the ratio $l_0^{\mathrm{S}}/l_0^{\mathrm{Fe}}$ almost exactly equal to unity. The steep nature of the power spectra of S and Fe for large $l$ ($|n_2|>5/3$) also matches characteristics of the power spectra of our models, confirming the absence of a turbulent cascade in 0509.

%More details on the fit procedure are provided in the Appendix.

%The other feature of interest in the PS is their slope at large $l$-values, or small scales. According to the fit, the PS of S and Fe distributions at small scales can be approximated as $C_l \propto l^{-2.1}$ and $C_l \propto l^{-2.5}$, respectively. This shows that the PS are too steep compared to a Kolmogorov cascade ($C_l \propto l^{-5/3}$). Thus, the PS for 0509 suggest a dynamical age $t_D\lesssim0.9$, since they haven't transitioned to Kolmogorov cascade yet (see Section~\ref{subsec:ps_models}).

\subsection{Trends in \los and \lofe of double detonation SNRs: implications for SNR 0509-67.5}
\label{subsec:model_vs_0509}

\begin{figure}
\centering
\includegraphics[width=0.45\textwidth]{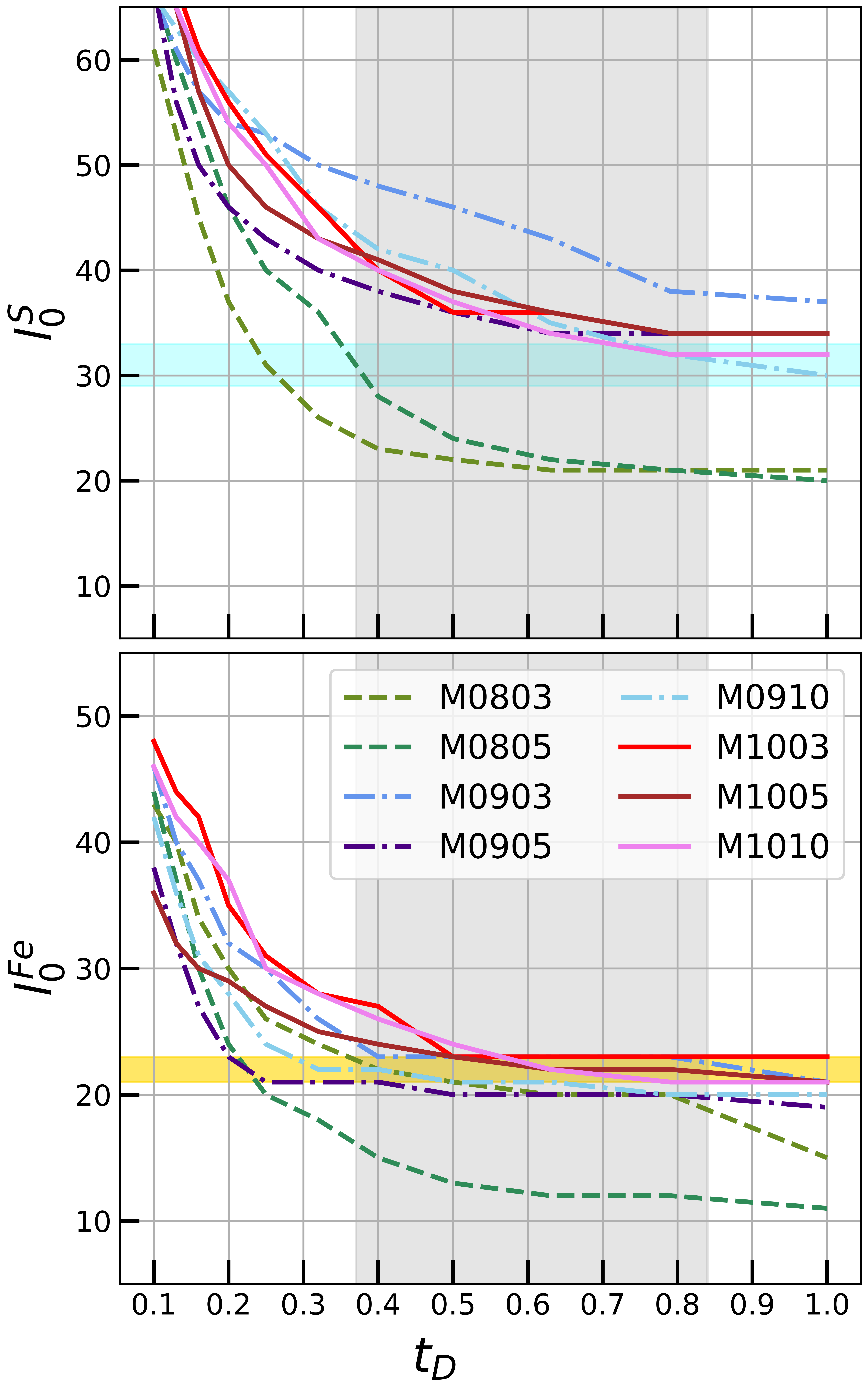}
\caption{Evolution of \los (\textit{top}) and \lofe (\textit{bottom}) with time for all DD SNR models. Results for models with core masses of $0.8M_{\odot}$, $0.9M_{\odot}$ and $1.0M_{\odot}$ have been plotted with dashed, dash-dotted, and solid lines respectively for ease of viewing. These plots have been overlaid with the range of \los ($=31\pm2$) and \lofe ($=22\pm1$) obtained from our fit for the power spectra of 0509, along with the likely range of dynamical age ($t_D$) of 0509.}
\label{fig:l0_vs_t}
\end{figure}

In Fig~\ref{fig:l0_vs_t}, we plot the time evolution of \los (top panel) and \lofe (bottom panel) for all of our models. The general trend of $l_0$ decreasing with time is observed. In addition, all $l_0$ values asymptote to a minimum at late times, as also noted by \cite{Mandal+2024ApJ}. We also overlay the panels with the most likely range of \los and \lofe for 0509 (as obtained in Section~\ref{subsec:ps_0509}), along with the possible range of dynamical age ($t_D$) of 0509, which has to be calculated using estimates of its age and Sedov time ($T_c$). The age of 0509 is estimated to be $400\pm120$ years from light echo modeling \citep{Rest+2005Nature}. \add{A range of values for the Sedov time are calculated using the ejecta masses and the explosion energies of our SNR models, as provided in Table~\ref{table:sn_models}. This calculation also requires a value for the ambient density, which is taken to be $0.4\,\mathrm{cm}^{-3}$, following \cite{Seitenzahl+2019PhRvL}. Using the range of age and Sedov time estimates for 0509, we find its dynamical age to be in the range $t_D=0.37-0.84$.}

%The Sedov time is calculated assuming an ejecta mass of $1M_{\odot}$, an explosion energy of $1.5\times10^{51}$ ergs and ambient number density of $0.4\,\mathrm{cm}^{-3}$, as suggested by \cite{Seitenzahl+2019PhRvL} on the basis of shock dynamics and emission measures observed in 0509. Using these values in Eqn.~\ref{eq:dynamical_age}, we obtain a range of $t_D=0.43-0.80$.

% While computing t_D, calculate largest possible range. The lower limit is calculated by taking lower limits on t and M and upper limits on E (M and E should match simultaneously). Calculate upper limit similarly.

% tD_lo = (400-120)/628 * (0.83/1.4)^(5/6) * (0.75)^(-1/2) * 0.4^(-1/3)
% tD_hi = (400+120)/628 * (1.05/1.4)^(5/6) * (1.42)^(-1/2) * 0.4^(-1/3)

The top panel of Fig.~\ref{fig:l0_vs_t} shows that the asymptotic value of \los at late times is dependent upon the core mass. Models with lower core mass (M0803 and 0805; dashed lines) exhibit a lower asymptotic value of \los in comparison to higher core mass models. In other words, lower core mass models tend to form larger IME clumps in comparison to their higher core mass counterparts. This trend is not monotonic, since the asymptotic value of \los for the model M0903 is found to be greater than the same for the M10xx models, but a stark difference is found between the models with core mass above and below $0.8M_{\odot}$. In contrast, \los isn't found to be sensitive to the mass of the He shell.

The bottom panel of Fig.~\ref{fig:l0_vs_t} shows that the asymptotic value of \lofe is lower than that of \los for each model, as expected. \add{It is also found that the asymptotic value of \lofe increases with increasing core mass.} Focusing on a particular value of core mass, one finds that the models with lower shell masses \add{tend to have a larger asymptotic value} of \lofe (compare the \lofe curves for M0803 and M0805, for example). Nevertheless, there is sufficient degeneracy in \lofe due to the varying core masses of our models that it would be difficult to constrain the shell mass unless the core mass was known accurately.
  
%The bottom panel of Fig.~\ref{fig:l0_vs_t} shows that the asymptotic value of \lofe is lower than that of \los for each model, as expected. At first glance, it appears that no clear trend exists for the asymptotic values of \lofe amongst the models. However, focusing on a particular value of core mass, one finds that the models with lower shell masses \add{tend to have a larger asymptotic value} of \lofe (compare the \lofe curves for M0803 and M0805, for example). Nevertheless, there is sufficient degeneracy in \lofe due to the varying core masses of our models that it would be difficult to constrain the shell mass unless the core mass was known accurately.

Comparing the values of \los in our models to those found for 0509, we find that models with mass $\leq0.8M_{\odot}$ are not consistent with 0509. The only models favored by the \los measurements are M0910 and M1010. The value of \los measured for 0509 also restricts us to $t_D\gtrsim0.7$, as can be seen from the top panel of Fig.~\ref{fig:l0_vs_t}. Looking at the range $0.70<t_D<0.84$ in the bottom panel of Fig.~\ref{fig:l0_vs_t}, we see that the measured value of \lofe in 0509 favors the M0903, M1003, M1005, and M1010 models. We therefore conclude that M1010 is the model in our suite that best reproduces the observed values of both \los and \lofe in 0509 simultaneously. \add{However, the late-time morphology of M0910 is also reasonably similar to that of 0509, with $l_0^{\mathrm{S}}=30\mbox{--}32$ and $l_0^{\mathrm{Fe}}=20$ for $t_D=0.70\mbox{--}0.84$. The predicted \los lies within the $1\sigma$ observational interval, while \lofe differs from the measured value for 0509 by $\approx2\sigma$. A discrepancy at this level does not constitute a statistically significant exclusion, but rather indicates that the model is disfavored relative to M1010. Models with thinner He shells are more strongly inconsistent with the observations.}

\add{Taken together, the model comparisons indicate that 0509 is best explained by a DD explosion of a sub-$\mathrm{M_{ch}}$ WD with a relatively massive CO core and He shell, as required by both the successful model (M1010) and the marginally consistent model (M0910). Interpreted within the parameter space sampled by our model grid, this corresponds approximately to CO core masses of $0.9–1.0M_{\odot}$ and He shell masses in the range $0.05M_{\odot}<M\lesssim0.1M_{\odot}$. A finer exploration of the CO core and He-shell mass parameter space would be required to determine tighter quantitative bounds.}

% We note that M0910 does reproduce values of \los and \lofe observed for 0509, but not at the same value of $t_D$. However, considering uncertainties in our numerical and observational analysis, we err on the side of caution and suggest that 0509 has a WD progenitor with a CO core mass of $0.9-1.0M_{\odot}$ and a He shell mass of $0.05M_{\odot}<M\lesssim0.1M_{\odot}$. We also note that our SNR models do not examine the range between \add{$0.05M_{\odot}<M<0.1M_{\odot}$} for the He shell mass due to limited availability of DD models, and hence cannot conclusively rule out models in this range.

We note that this constraint holds even if one estimates $t_D$ for 0509 assuming a near-$\mathrm{M_{ch}}$ WD progenitor, unlike our calculation. For instance, \cite{Arunachalam+2022ApJ} favor a near-$\mathrm{M_{ch}}$ origin for 0509, on the basis of forward shock kinematics measured using Hubble H$\alpha$ observations. Assuming $E_{51}=1.4$, they obtain $M=2.02\pm0.85M_{\odot}$, $\mathrm{log\,}\rho_0=-24.23\pm0.09$, and an age of $t=329.7\pm17.3$ years for 0509. Alternatively, holding $M$ fixed to $1.4M_{\odot}$, they estimate probable ranges $E_{51}=1.30\pm0.41$, $\mathrm{log\,}\rho_0=-24.20\pm0.20$, and an age of $t=317.19\pm12.95$ years. Using these values, we obtain a range of $t_D=0.65-0.95$. As Fig.~\ref{fig:l0_vs_t} shows, this range of $t_D$ still favors the M1010 model when the values of both \los and \lofe are considered. 

The M1010 model has an ejecta mass of $1.1M_{\odot}$ and kinetic energy $\approx1.5\times10^{51}\mathrm{\,ergs}$. Also taking the age of 0509 to be $400\pm120\mathrm{\,years}$ \citep{Rest+2005Nature}, we can compute a range of dynamical age for 0509 using Eqn.~\ref{eq:dynamical_age}:

\begin{equation}
    \label{eq:tD_0509}
    \begin{split}
        t_D &= \left(\frac{400\pm120\mathrm{yrs}}{628\mathrm{yrs}}\right)\left(\frac{1.1M_{\odot}}{1.4M_{\odot}}\right)^{5/6} 1.5^{-1/2}n_0^{-1/3} \\
        &= (0.43\pm0.13) n_0^{-1/3}.  \\
    \end{split}
\end{equation}

\add{We can estimate the number density of the AM currently being encountered by the forward shock in 0509, by comparing Eqn.~\ref{eq:tD_0509} to the range $0.70<t_D<0.84$ predicted by our analysis. This yields a range $0.05<n_0<0.5$ (in units of $\mathrm{cm^{-3}}$) for the AM density around 0509 today.} This is a rather broad range, but is consistent with all previous estimates for the AM density around 0509 \citep{Seitenzahl+2019PhRvL,Arunachalam+2022ApJ}.

\add{An additional, independent test of the models is provided by the physical size of 0509, which has been constrained to $\approx 3.6$ pc \citep{Hovey+2015ApJ,Arunachalam+2022ApJ}. Using the analytical framework of \cite{Truelove+1999ApJS}, \cite{Seitenzahl+2019PhRvL} demonstrated that an ejecta mass of $1M_{\odot}$, explosion energy of $1.5\times10^{51}$ ergs, and AM number density of $0.4\,\mathrm{cm^{-3}}$ can reproduce this observed size. For the M1010 model (with mass and energy as listed in Table~\ref{table:sn_models}), the forward-shock radius is 3.45 pc at $t_D=0.7$, in good agreement with the measured value. At later dynamical times (e.g., $t_D=0.8$), the model radius exceeds the observed size. We therefore find that the radius constraint for 0509 is fully compatible with the conclusions drawn from the substructure analysis. We stress, however, that this agreement should be interpreted as a broad consistency check rather than evidence that the M1010 model at $t_D=0.7$ uniquely or definitively describes 0509.}

%We stress, however, that this agreement should be interpreted as a broad consistency check rather than evidence that the M1010 model at $t_D=0.7$ uniquely or definitively describes 0509. A fully self-consistent model must simultaneously reproduce additional observational constraints, including detailed ejecta structure and abundance distributions. The radius comparison simply demonstrates that the dynamical scale implied by the substructure analysis is not in conflict with the independently measured angular size of the remnant.

\section{Discussion}    \label{sec:discussion}

\add{In this work, we delve deeper into the study by \cite{Mandal+2025ApJ} on the identification of the explosion mechanism and progenitor properties of Type Ia SNRs. Using numerical models of Type Ia SNRs that exploded via the double detonation (DD) mechanism, we compute the typical size of turbulent substructures in such SNRs. In addition to confirming that iron-dominated substructures are considerably larger than sulfur-dominated substructures in DD SNe \citep[as found by][]{Mandal+2025ApJ}, we show that these substructure sizes are sensitive to the carbon-oxygen core mass and the helium shell mass of the progenitor WD.}

We apply these results to the substructure size distributions in \add{the SNR 0509-67.5}, computed from high resolution MUSE observations. We infer spherical wavenumbers corresponding to the typical sulfur-dominated and iron-dominated substructures in SNR 0509-67.5 to be $l_0^{\mathrm{S}} = 31 \pm 2$ and $l_0^{\mathrm{Fe}} = 22 \pm 1$, respectively. These values correspond to characteristic substructure sizes of approximately $\approx5\%$ (sulfur-dominated) and $\approx7\%$ (iron-dominated) of the remnant radius (see section~\ref{subsec:image_anly} for discussion on conversion from angular harmonics to length scales). As mentioned earlier, this is a key signature of a sub-$\mathrm{M_{ch}}$ WD progenitor for Type Ia SNRs. \add{Our study thus provides novel and independent evidence to the suggestion that SNR 0509 originated from the double detonation of a sub-$\mathrm{M_{ch}}$ CO WD \citep{Das+2025NatAs}. In addition, our suite of SNR models based on DD SN models by \cite{Gronow+2021AandA} spanning a range of WD core mass and shell mass helps us constrain the core and shell masses of the progenitor of SNR 0509-67.5.} The progenitor is found to be the most consistent with a WD of core mass and shell mass of $1M_{\odot}$ and $0.05M_{\odot}<M\lesssim0.1M_{\odot}$, respectively. The total ejecta mass and energy ($\approx1.5\times10^{51}\mathrm{\,ergs}$) of this model is found to be in good agreement with the prediction of \cite{Seitenzahl+2019PhRvL}.

As noted in Section~\ref{sec:intro}, 0509 is also associated with a 1991T-like event based on light echo reconstruction of spectra \citep{Rest+2008ApJ}. These events are typically overluminous and exhibit slow photometric decline rates. The M1010 model considered here shows broadly similar properties, with high maximum absolute bolometric magnitude ($M_{\mathrm{bol,max}}\sim-19.4$) and relatively slow decline rates ($\Delta m_{15,\mathrm{bol}} \sim 0.96$) along some viewing directions \citep[see Fig. 6 of][]{Gronow+2021AandA}. Despite this qualitative similarity, detailed light curve calculations indicate that this model does not reproduce the observed light curve of SN~1991T \citep{Collins+2022MNRAS}. Nevertheless, recent non-LTE radiative transfer calculations suggest that double detonation models may be capable of producing light curves and spectra consistent with both normal and overluminous Type Ia SNe \citep{Shen+2021ApJ,Collins+2025MNRAS}. In this context, \cite{O'brien2024} argue that 1991T-like SNe may represent an extension of the normal Type~Ia population, differing primarily in elemental abundances and ionization structure.

\add{Observations have also begun to reveal individual 1991T-like events with signatures potentially consistent with helium shell detonations. For example, SN~2022joj shows shallow Si~II absorption features characteristic of the 1991T subclass and has been interpreted as a possible helium-shell double detonation event \citep{Liu+2023ApJ}. Modeling of this object suggests a helium shell mass in the range $\simeq0.04$--$0.1\,M_{\odot}$, comparable to the shell masses inferred in this work. Another example is SN~2020eyj \citep{Kool2023}, which initially exhibited a typical 1991T-like spectrum and light curve before later showing evidence for interaction with helium-rich circumstellar material (CSM). Such behavior may be related to the broader connection between 1991T-like SNe and SNe~Ia-CSM noted by \cite{Leloudas+2015AandA}, who argue that some 1991T-like events may arise in systems that produce dense CSM through mass loss from a non-degenerate companion. It is important to note that CSM interaction in these systems typically manifests within the first $\sim100$ days of the SN evolution and is not necessarily expected in the remnant phase. The relatively low ambient densities inferred for SNR~0509 therefore do not conflict with the possibility of early-time CSM interaction in some 1991T-like events.}

%Recent observations of SN 2022joj \citep{Liu+2023ApJ,Gonzalez2024} and 2020eyj \citep{Kool2023} have also pointed towards the possible origin of 1991T-likes from a double detonation event. An alternative explanation for the high luminosities of 1991T-like SNe invokes the presence of dense circumstellar medium (CSM) in the \add{immediate} vicinity of such SNe, which could enhance radiative output by converting kinetic energy into radiation. This has been proposed by \cite{Leloudas+2015AandA}, who find an association between 1991T-like SNe and SNe Type Ia-CSM (Type Ia SNe interacting with dense CSM). They suggest that 1991T-like SNe are likely born in single degenerate systems generating dense CSM due to mass loss from the main sequence companion.

\add{Recent observations of SNR~0509 itself may indicate a helium-enriched environment. Das et al.~(2026, in review) report elevated He/H abundance ratios in Balmer-dominated filaments of 0509, pointing to a helium-rich progenitor environment. While the origin of this enrichment remains uncertain (see Das et al.~2026 for more discussion of possible progenitor scenarios), it is broadly consistent with scenarios involving a white dwarf with a substantial helium layer prior to explosion, as inferred from our analysis.}

As of now, no explosion models can completely explain the observed features of 1991T-like SNe due to inherent multidimensional nature of the explosion \citep{Pakmor2024}. Given the growing line of evidence that SNR 0509-67.5 originated from a double detonation \citep[][and this work]{Seitenzahl+2019PhRvL,Das+2025NatAs}, we encourage future works on the double detonation model to investigate the effects of multidimensionality and non-LTE on their lightcurves and spectra, as well as study the connection between 1991T-like events and double detonation SN models with dense CSM.

\acknowledgments
 
We thank the anonymous referee, whose careful scrutiny significantly improved this manuscript. This work made use of the Heidelberg Supernova Model Archive (HESMA), https://hesma.h-its.org. Numerical calculations were performed on the Rivanna computing cluster at University of Virginia.  P.G. acknowledges support from the Maryland Space Grant Consortium and NASA.

\vspace{8mm}

\software{\sprout\, \citep{Mandal+2023_sprout},  
    VisIt \citep{HPV:VisIt}, 
    SHTOOLS \citep{SHTOOLS},
    NumPy \citep{numpy},
    Matplotlib \citep{matplotlib},
    Astropy.
}

\bibliographystyle{apj} 
\typeout{}
\bibliography{smbib}

\end{document}